\documentclass[useAMS,usenatbib]{mn2e}

\usepackage{rotating}
\usepackage{lscape}
\usepackage{graphicx}
\usepackage{amssymb}
\usepackage{color}
\usepackage{soul}
\usepackage{multirow}

\title[PCA and Radiative Transfer of Spitzer IRS spectra of ULIRGs]
  {Principal Component Analysis and Radiative Transfer modelling of  \textit{Spitzer} IRS Spectra of Ultra Luminous Infrared Galaxies}
\author[P.D. Hurley et al.]{P.D.~Hurley,$^1$\thanks{Email: P.D.Hurley@sussex.ac.uk} S.~Oliver,$^1$ D.~Farrah,$^1$,$^3$ L.~Wang,$^1$ A.~Efstathiou$^2$ \\
$^1$Astronomy Centre, Department of Physics and Astronomy, University of Sussex, Falmer, Brighton BN1 9QH, UK\\
$^2$School of Sciences, European University Cyprus, Diogenes St, Engomi, 1516 Nicosia, Cyprus\\
$^3$Virginia Polytechnic Institute \& State University, Department of Physics,MC 0435, 910 Drillfield Drive, Blacksburg, VA 24061\\ }

\date{Released 2002 Xxxxx XX}

\pagerange{\pageref{firstpage}--\pageref{lastpage}} \pubyear{2002}

\def\LaTeX{L\kern-.36em\raise.3ex\hbox{a}\kern-.15em
    T\kern-.1667em\lower.7ex\hbox{E}\kern-.125emX}

\graphicspath{{./newerPLOTS/}}
\begin{document}

\label{firstpage}
\maketitle
\begin{abstract}
The mid-infrared spectra of ultraluminous infrared galaxies (ULIRGs) contain a
variety of spectral features that can be used as diagnostics to characterise the
spectra.  However, such diagnostics are biased by our prior prejudices on the origin of the features. Moreover, by using only part of the spectrum they do not utilise the full information content of the spectra.
Blind statistical techniques such as principal component analysis (PCA) consider the whole spectrum, find correlated features and separate them out into distinct components. 

We further investigate the principal components (PCs) of ULIRGs derived in
\cite{LingyuPCA}. We quantitatively show that five PCs is optimal for describing the IRS spectra.  These five components (PC1-PC5) and the mean spectrum provide a template basis set that reproduces spectra of all $z<0.35$ ULIRGs within the noise.
For comparison, the spectra are also modelled with a combination of radiative transfer models of both starbursts
and the dusty torus surrounding active galactic nuclei. The five PCs typically provide better fits than the models. We argue that the radiative transfer
models require a colder dust component and have difficulty in
modelling strong PAH features.

Aided by the models we also interpret the physical processes that the principal components represent. The third
principal component is shown to indicate the nature of the dominant power source, while PC1 is related to the inclination of the AGN torus. 

Finally, we use the 5 PCs to define a new classification scheme using 5D Gaussian mixtures modelling and trained on widely used optical classifications. The five PCs, average spectra for the four classifications and the code to classify objects are made available at: \url{http://www.phys.susx.ac.uk/~pdh21/PCA/}.
\end{abstract}
\section{Introduction}
Ultraluminous Infrared Galaxies (ULIRGs) are galaxies whose rest-frame infrared luminosities, $L_{8-1000\mathrm{\mu m}}$, exceed $10^{12}\mathrm{L_{\odot}}$. Although ULIRGs were first discovered using ground based photometry in the 1970s \citep{Rieke:1972cy}, the \emph{IRAS} survey transformed our understanding by observing the objects in much larger numbers \citep{Soifer:1984nm}. Most have high star-formation rates ($SFR > 100\mathrm{M_{\bigodot}yr^{-1}}$), while around half also contain an embedded Active Galactic Nucleus (AGN). 

ULIRGs are rare in the local Universe, with less than fifty at $z \lesssim 0.1$, but the associated luminosity function shows strong, positive evolution with redshift \citep[e.g.][]{Sanders:1999py}, resulting in several hundred ULIRGs per square degree at $z > 1$  \citep{Rowan-Robinson:1997, Barger:1998jm, Hughes:1998id, Eales:2000bs, Fox:2002mc, Floch:2005pe}. The increase in number density with redshift and their associated high SFR means ULIRGs make a significant contribution to the history of star formation at high $z$.

The mid to far infrared luminosity of ULIRGs is a result of dust and gas reprocessing the optical and UV radiation emitted by stars and/or AGN. Obtaining spectroscopy for the mid-infrared part of the spectrum became possible with instruments such as the \emph{Infrared Space Observatory} (\emph{ISO}; \cite{Kessler:1996nl}), and the Infrared Spectrograph (\emph{IRS}; \cite{Houck:2004pi}) on the \emph{Spitzer Space Telescope} \citep{Werner:2004gm}. The ULIRG spectra from these instruments contain a wealth of spectral features. These include the emission lines from broad polycyclic aromatic hydrocarbons (PAHs), which are strong in starforming regions, but absent in AGN dominated sources \citep{Moorwood:1986, Roche:1991}. A prominent [Ne V] 14.3 $\mathrm{\mu m}$ fine structure line indicates the presence of an AGN, while the silicate features at 9.7 and 18 $\mathrm{\mu m}$ probe source geometry \citep{Imanishi:2007ux}. 

Combinations of the PAH emission lines, mid-infrared fine-structure lines and silicate features have been used as diagnostics for characterising the power source behind the ULIRGs \citep{Genzel:1998, Rigopoulou:1999,Spoon:2007uq, Farrah:2007fe, Farrah:2008uk,Farrah:2009}. There are however problems associated with these diagnostic tools, such as the separation of emission lines from both the continuum and underlying PAH features, the mixture of neighbouring features and different diagnostics giving conflicting estimates. They also only focus on small parts of the spectrum, disregarding the information contained in the remainder.

Larger regions of the spectrum can be investigated with the multivariate statistic, Principal Component Analysis (PCA). PCA has been used for spectral classification for optical galaxies \citep[e.g.][]{Connolly:1995,Bromley:1998}. \cite{LingyuPCA} carried out PCA on the $\mathit{IRS}$ spectra of 119 local ULIRGs. They argued, qualitatively, that only 4 principal components (PCs) were needed to reproduce the variance in the ULIRG spectra. They also proposed that the contribution from each PC had some underlying physical interpretation. Examination of the first four PCs, and comparisons to the diagnostics employed by \cite{Spoon:2007uq} and  \cite{Nardini:2009kr} suggested that PC1 constrains the dust temperature and geometry of the distribution of source and dust, while PC2 and PC3 determine the amount of star formation. The fourth PC is important for Seyfert Type 2 galaxies, and is hence a possible indicator of an unobscured AGN. 

In this paper we extend \cite{LingyuPCA} by quantitatively investigating how many PCs are needed to explain the variation in the spectra and compare the PC reconstructions to fits provided by a suite of radiative transfer models. We investigate what information the radiative transfer models are missing. We also re-examine what physical properties are behind the PCs, by investigating the relationship between the physical parameters of models and the contributions from different PCs. Finally, we introduce a new classification scheme using 5D Gaussian mixtures modelling and trained with optical classifications. Section \ref{Data} gives an overview of the data and Section \ref{PCAspace} a brief description of PCA. Section \ref{radmodels} will review the radiative transfer models being applied, and Section \ref{results} will present the results. Conclusions will be presented in Section \ref{conclusions}. We assume a spatially flat cosmology with $H_{0} = 70 \mathrm{km s^{-1} Mpc^{-1}}, \Omega = 1$, and $\Omega_{m} = 0.3.$


%
\section{The Data}
\label{Data}
This paper uses the same sample of mid-infrared spectra as \cite{LingyuPCA}. We summarise their selection criteria here. The ULIRGs were observed as part of the IRS Guaranteed Time program \citep{Armus:2007, Farrah:2007fe, Spoon:2007uq} and those observed by \cite{Imanishi:2007ux}. An upper redshift cut of z = 0.35 was applied to ensure we sample approximately the same wavelength range for each object. A further eight objects were removed as they have poor-quality data in the longer-wavelength IRS module. In total, there are 119 objects in the sample.

\section{Principal Component Analysis (PCA)}
\label{PCAspace}
PCA works by determining the eigenvectors from the covariance matrix of a given dataset. For 119 spectra, each with 180 wavelength points, the 180 by 180 covariance matrix quantifies the correlation between each spectral point. The eigenvectors of the matrix can be thought of as spectral components that can be linearly combined to reconstruct each object in the sample.

Any spectrum can be linearly decomposed by projecting it onto the principal components defined by the 119 ULIRG sample. This allows each spectrum to be described by the contribution from each PC. These contributions define co-ordinates in a multidimensional space which we refer to as PCA space. 
\section{Radiative transfer models}
\label{radmodels}

To compare with the fits provided by the principal components, we have carried out a minimum chi squared search for linear combinations of a grid of starburst models described in \cite{Siebenmorgen:2007bc} and grid of AGN dusty torus models of \cite{Efstathiou:1995ge}. The libraries contain 5948 and 2109 SEDS respectively and we have considered linear combinations of each AGN and starbust SED, giving us many models to search over.

\subsection{Starburst Models}

We use the \cite{Siebenmorgen:2007bc} starburst models. The models presented by \cite{Siebenmorgen:2007bc} have been described as 'hot spot' starbursts. OB stars are assumed to be surrounded by dense clouds (the hot spots) and other stars, such as old bulge stars or massive stars are dispersed in the diffuse medium. It is the hot spots that contribute to the mid infrared part of the spectrum. The outer radius of these environments is determined by the condition of equal heating of the dust by the OB stars in the centre and the interstellar radiation field. 

Both stellar groups are treated as continuously distributed sources, and the number density of both types of stars, falls off as $r^{-1}$.

The parameters of these models include the starburst radius, $R$;  ratio of the luminosity of OB stars with hot spots to total luminosity, $f_{OB}$; the total luminosity of the starburst, $L_{SB}$; total extinction from the outer radius of the galactic nucleus to its centre, $A_{v}$; and dust density of the hot spot environment, $\rho_{HS}$, corresponding to hydrogen number densities ($n_{HS}$) and assuming a gas to dust ratio of 150. The parameter ranges can be found in Table \ref{Paramtable}. In total, the library contains 5948 SEDs.

\begin{table}
\centering
\begin{tabular}{|l|l|}
\hline
Parameter & Range \\
\hline
$R$ (kpc) & 0.35, 1 and 3 \\
$f_{OB}$ & 0.4, 0.6 and 0.9 \\
$L_{SB}$ ($\mathrm{L_{\odot}}$) & $10^{10} $ to $10^{14}$ in steps of 0.1 dex\\
$A_{v}$ (Mag) & 2.2, 4.5, 7, 9, 18, 35, 70 and 120 \\
$n_{hs}$  ($\mathrm{cm^{-3}}$) & $10^{2},10^{3},2.5\cdot10^{3},5\cdot10^{3},7.5\cdot10^{3},10^{4}$ \\
\hline
\end{tabular}
\caption{Parameter values and ranges for the starburst models.}
\label{Paramtable}
\end{table}

\subsection{AGN torus models}
This paper uses the AGN tapered disc models of \cite{Efstathiou:1995ge}. The tapered disc models, in combination with the starburst models of \cite{Efstathiou:2000zc}, have been successful
in fitting the spectral energy distributions of ultraluminous
infrared galaxies \citep{Farrah:2003ge}, hyperluminous infrared
galaxies \citep{Farrah:2002, Verma:2002, Efstathiou:2006}
submillimeter galaxies \citep{Efstathiou:2009bd},
and active galaxies \citep{Alexander:1999,Efstathiou:2005,Farrah:2012,Ruiz:2001}. The torus is modelled as a disc, whose thickness increases with distance from the central source but tapers off in the outer regions of the torus. The dust density is distributed smoothly within the disc and follows a $r^{-1}$ relation, with r being radius. The parameters for the AGN torus model are: ultraviolet equatorial optical depth to the centre of the torus, $\tau$; the opening angle of the torus, $\Theta$; the ratio of inner to outer radius of the torus, $r_{\mathrm{in}}/r_{\mathrm{out}}$; and the viewing angle, $\theta$. In total, there are 2109 AGN SEDs.

\begin{table}
\centering
\begin{tabular}{|l|l|}

\hline
Parameter & Range \\
\hline
$\tau$ & 500, 750, 1000, 1250 \\
$\Theta$ (degrees)& 30, 45, 60 \\
$r_{\mathrm{in}}/r_{\mathrm{out}}$ & 20, 60, 100 \\
$\theta$ (degrees) & 0 to $90$ with either\\
&40 or 75 divisions (depending on $r_{\mathrm{in}}/r_{\mathrm{out}}$)\\
\hline
\end{tabular}
\caption{Parameter values and ranges for the AGN models.}
\label{Paramtable_AGN}
\end{table}
\subsection{The fitting procedure}
\label{subsec:chi}

We have considered all linear combinations of a starburst and AGN model when fitting the observed spectra of the 119 ULIRG sample.  We use the wavelength grid of the starburst models, and the lower resolution AGN models are interpolated onto the same grid. The smoothness of the AGN models, makes the interpolation justifiable. The radiative transfer models lack molecular hydrogen emission so we mask out regions of the spectrum where molecular hydrogen features occur ( i.e. $9.46-9.86, 12.08-12.48$ and $16.83-17.23 \mathrm{\mu m}$).

The wavelength resolution of the PCs is higher than the starburst model resolution. For proper comparison to the fits, and to allow decomposition of the models into PCA space, we have re-derived the principal components for the ULIRG sample at the resolution of the starburst models. There is no significant change in the shape of components. We also note that the sign of the PC contributions for each object, remains the same and the change in magnitude of the PC contributions is not significant in comparison to the spread of contributions for the sample. 

To remain consistent with the analysis of \cite{LingyuPCA}, the models are normalised so that the mean flux over the whole wavelength range is unity.  

We then carry out a linear least squares fit for each combination of starburst and AGN model, with the condition that the fit parameters are positive (i.e. to eliminate the possibility of a negative amount of starburst or AGN). Model comparison is then carried out via minimum chi squared ($\chi ^{2}$).

We assumed a minimum of 5\% flux error for each spectral bin of the IRS spectra, which is consistent with the observed variations between individual nod positions on the IRS as described in Chapter 7 of the IRS Instrument Handbook\footnote{http://irsa.ipac.caltech.edu/data/SPITZER/docs/irs/}.

\section{Results}
\label{results}
\subsection{Optimum number of components}
\begin{figure}
\includegraphics[width=8.5cm]{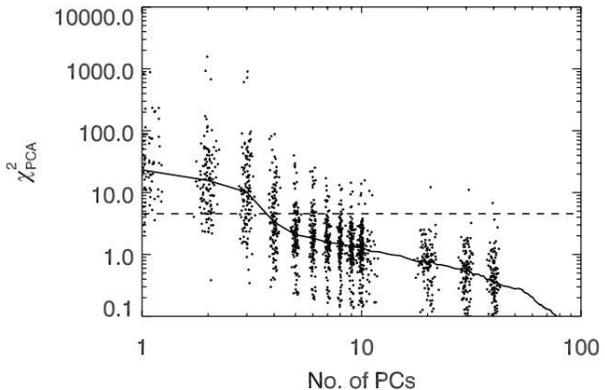}
\caption{The median variation of the $\chi _{\nu}^{2}$ for the PC reconstruction as the number of components used in the reconstruction are increased. The dashed line indicates the median $\chi _{\nu}^{2}$ for the radiative transfer model fits. For PC reconstructions using up to 10 PCs (and 20, 30, 40) we also plot the $\chi _{\nu}^{2}$ for every object (offset for clarity)}\label{chivsno.pcs}
\end{figure}

\begin{figure}
\includegraphics[width=8.5cm]{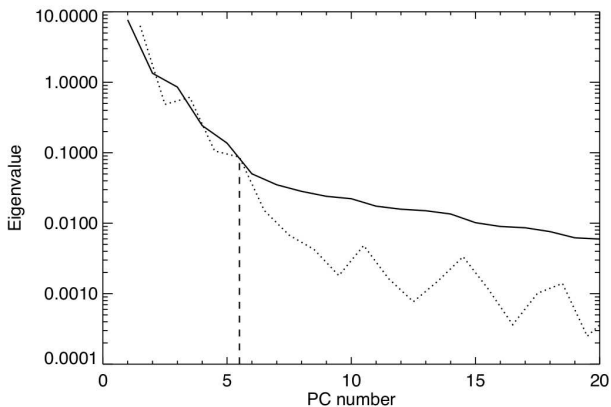}
\caption{The eigenvalues (solid line) and difference in eigenvalues (dotted line) for the PCs. The eigenvalues quantify the variance associated with each PC, and are a measure of importance. The difference between eigenvalues drops dramatically for the first few PCs, but levels off beyond 5 (indicated by the dashed line). We therefore argue that 5 PCs is a more suitable number than the 4 used in \protect\cite{LingyuPCA}.}
\label{scree}
\end{figure}

We first investigate how many PCs are needed to describe the ULIRG sample. \cite{LingyuPCA} did not quantitatively show whether 4 PCs were sufficient. Using the PCs re-derived at the lower resolution described in Section \ref{subsec:chi}, we have investigated how many PCs are needed to accurately reconstruct the IRS spectra of all 119 ULIRGs in the sample. For each spectrum, we quantify the goodness of reconstruction with the reduced chi squared statistic $\chi _{\nu}^{2}$, where the number of degrees of freedom is equal to the number of wavelength points minus the number of PCs used in the reconstruction.

\begin{figure*}
\includegraphics[width=16cm]{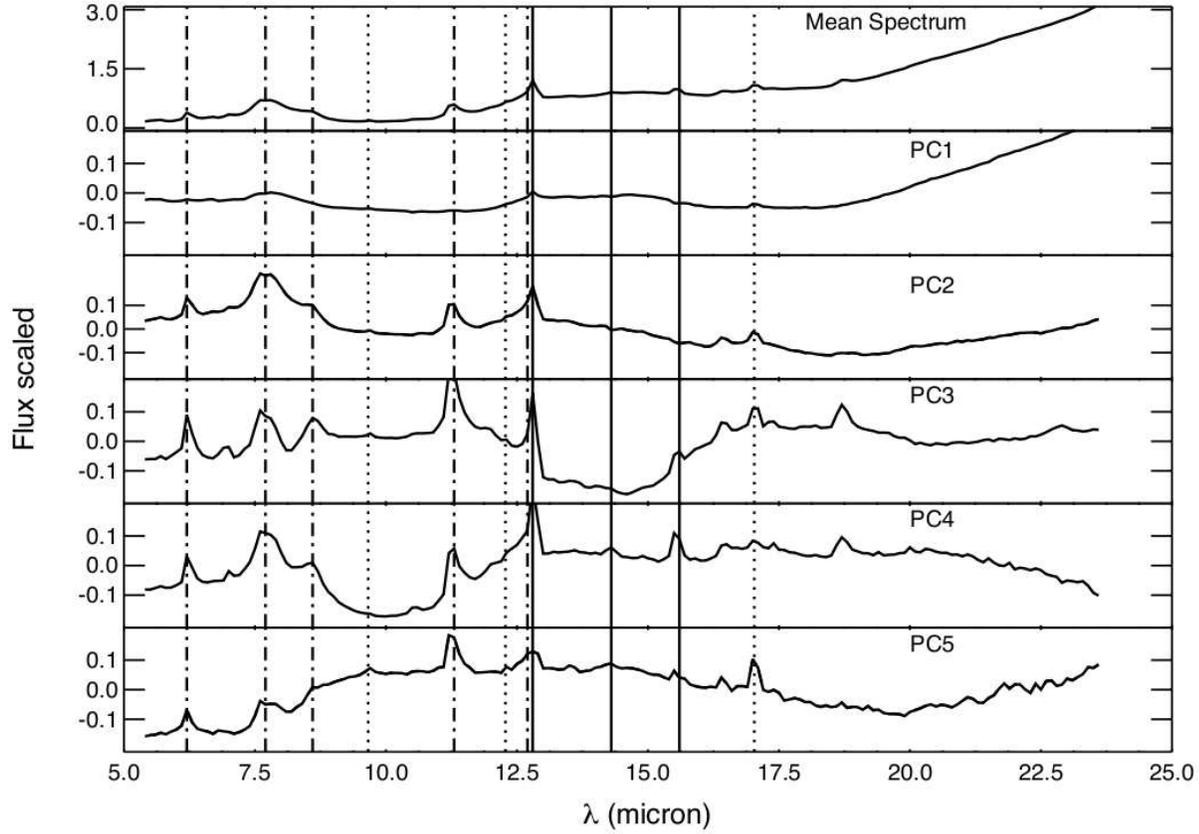}
\caption{The mean spectrum and principal components for the sample of ULIRGs. The dot-dashed vertical lines mark the central location of the 6.2,7.7,8.6,11.2 and 12.7 $\mathrm{\mu m}$ PAH emission lines. The dotted lines indicate the location of the molecular hydrogen lines at 9.66, 12.28 and 17.03 $\mathrm{\mu m}$. The solid vertical lines indicate the position of the neon fine-structure lines, [Ne II] 12.8, [Ne v] 14.3 and [Ne III] 15.6 $\mathrm{\mu m}$.} 
\label{PCs_low}
\end{figure*}

Figure \ref{chivsno.pcs} shows that as we increase the number of PCs used in the reconstruction, the median $\chi _{\nu}^{2}$ value for the sample decreases. We have plotted the $\chi _{\nu}^{2}$ for each individual object for reconstructions using up to ten PCs and the median $\chi _{\nu}^{2}$ value obtained by fitting the ULIRGs with the radiative transfer models as described in Section \ref{subsec:chi}. Ten PCs would appear to be the optimal number i.e. where $\chi _{\nu}^{2}=1$. We find that four PCs (assumed by \cite{LingyuPCA}) give a median $\chi _{\nu}^{2}$ of 3.3, while adding a fifth component substantially decreases the median $\chi _{\nu}^{2}$ to 2.1. The use of six and seven PCs only reduces the median $\chi _{\nu}^{2}$ to 1.8 and 1.6 respectively. 

The eigenvalues associated with each PC are a measure of the variance each PC accounts for and provide an alternative method to determine the optimum number of components. In Figure \ref{scree}, we plot the eigenvalues and difference in eigenvalues for the PCs. The general trend indicates the difference between eigenvalues significantly decreases with each component. The exception to the rule occurs between the 3rd-4th component and the 5th-6th component where the difference between eigenvalues is larger than the trend. We associate this larger than expected difference as in indication that the previous component captures significantly more information than the next. This suggests that the third and fifth PC are substantially more important than the fourth and sixth respectively. Beyond the sixth PC, the trend flattens out, indicating most of the variation related to structure has been captured. Overall, Figures \ref{chivsno.pcs} and \ref{scree} do not definitively indicate the optimum number of PCs. However, we argue that the reduction in $\chi _{\nu}^{2}$ to 2.1 and difference in eigenvalue between the fifth and sixth PC, indicates that five PCs rather than the four PCs used by \cite{LingyuPCA}, strike a better balance of providing a small basis set of templates, whilst adequately describing the spectra.

The fifth component was not discussed in \cite{LingyuPCA} and so we now show this component, compared to the original four. The mean spectra of the 119 ULIRGs and the 5 components can be seen in Figure \ref{PCs_low}. There are a number of spectral features in this fifth component, most notably the 6.2, 11.2 and 12.7 $\mathrm{\mu m}$ PAH emission lines as well as the molecular hydrogen emission line at 17.03 $\mathrm{\mu m}$. The 6.2 $\mathrm{\mu m}$ emission feature has negative flux, while the 11.2 and 12.7 PAH lines are both positive. Overall, the fifth component does not contain any new features that were not seen in the previous components. Its role appears to be in altering the ratios of existing features.

\subsection{Analysis of the radiative transfer models}
\begin{figure}
\centering
\begin{tabular}{cc}
\includegraphics[width=8.5cm]{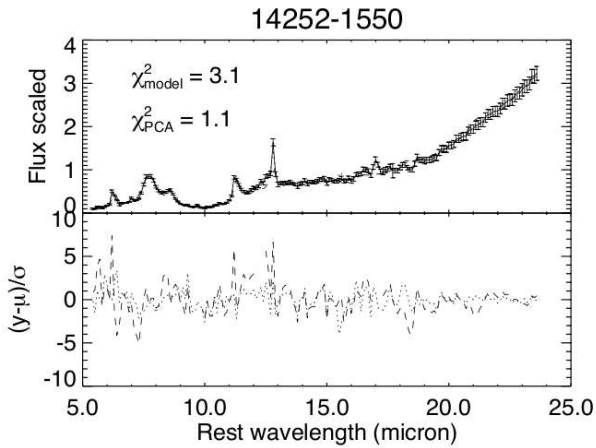}&
\end{tabular}
\caption{An example of our fit with 14252-1550. The radiative transfer model is plotted with a dashed line, and the principal component reconstruction with 5 PCs is shown with a dotted line. The residual over error is also shown to indicate where either technique may be failing}\label{specexamps}
\end{figure}

We now investigate whether the radiative transfer models discussed in Section \ref{radmodels} are capable of modelling the spectra. An example of the fit produced by 5 PCs and the radiative transfer models can be seen in Figure \ref{specexamps}.

\begin{figure}
\includegraphics[width=8.5cm]{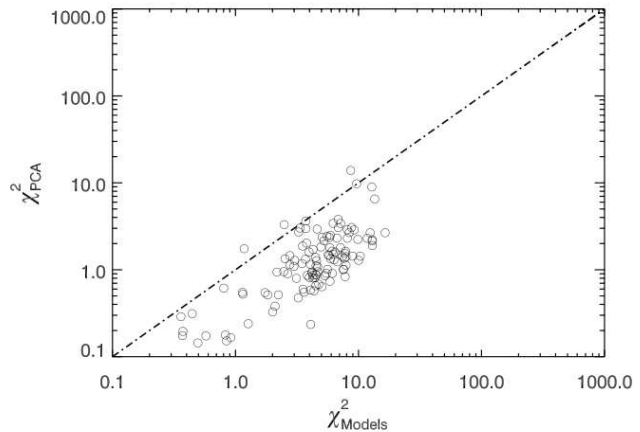}
\caption{The $\chi _{\nu}^{2}$ values for each object in the sample for both radiative transfer model fits and the 5 PC reconstruction. Most objects do better with the PCs.}\label{chivschi}
\end{figure}

\begin{figure}
\includegraphics[width=8.5cm]{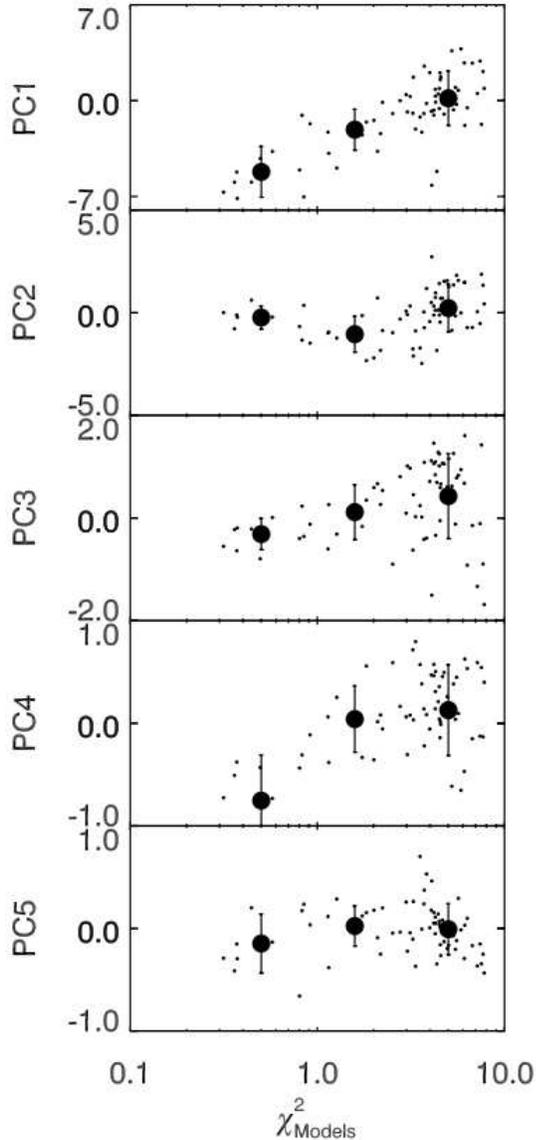}
\caption{The contributions made by each PC against the $\chi _{\nu}^{2}$ from the model fits. Only objects with a 5 PC fit of $\chi _{\nu}^{2} \leq 3$ have been plotted. The mean and one sigma dispersion for three bins are overplotted as filled circles and errorbars}\label{pcschi}
\end{figure}

We now compare all the $\chi _{\nu}^{2}$ for reconstructions using 5 PCs with the $\chi _{\nu}^{2}$ for our radiative transfer model fits. Figure \ref{chivschi} shows the distribution of the reduced chi squared values for both the 5 PC reconstructions and the radiative transfer model fits, for all ULIRGs in the sample. A 5 component reconstruction fits the spectra better, on average, than the radiative transfer models.

We have shown that 5 PCs can explain the sample of ULIRGs better than the radiative transfer models, but the two are not competing methodologies. The PCs will always do better than the models as they are derived from the data and the number of PCs is increased until the reproduction of the spectra is good. They represent an extraction of most of the important information from the spectra. Radiative transfer models are used to give us physical information of objects. However, Figure \ref{chivschi} indicates that the ULIRGs are not modelled well on average by the radiative transfer models. 

By comparing the models to the PCs, we can investigate what information is in the PCs that is not in the models. Figure \ref{pcschi} shows the contributions made by the five PCs, as a function of $\chi _{\nu}^{2}$ for the model fits. We only plot objects that have a reasonable $\chi _{\nu}^{2}$ for the 5 PC reconstruction i.e. a $\chi _{\nu}^{2} \leq 3$. We also bin the model $\chi _{\nu}^{2}$ values into three bins. The mean and one sigma dispersion are overplotted as filled circles and errorbars.

The general increase of $\chi _{\nu Models}^{2}$ in Figure \ref{pcschi} shows that models tend to do worse when the objects have a large, positive contribution from PC1. \cite{LingyuPCA} suggested a large, positive contribution from PC1 indicated colder dust. Our results suggest that objects with colder dust are not well modelled by the AGN and starburst component models. The increase in dispersion with $\chi _{\nu Models}^{2}$ for PC2 and PC3 indicates models do worse when there is a large, absolute contribution from PC2 and PC3. PC2 and PC3 relate to strong spectral lines, which would indicate that the models have problems with constraining the strength of spectral lines. Lower values of $\chi _{\nu Models}^{2}$ appear to occur when objects have negative values of PC4, but as the $\chi _{\nu Models}^{2}$ values increase beyond 2, there appears to be little change in PC4. A negative contribution in PC4 would suppress emission features, indicating that models are again inadequate in modelling spectral features. There appears to be little change of PC5 contribution with $\chi _{\nu Models}^{2}$.

\begin{figure}
\includegraphics[width=8.5cm]{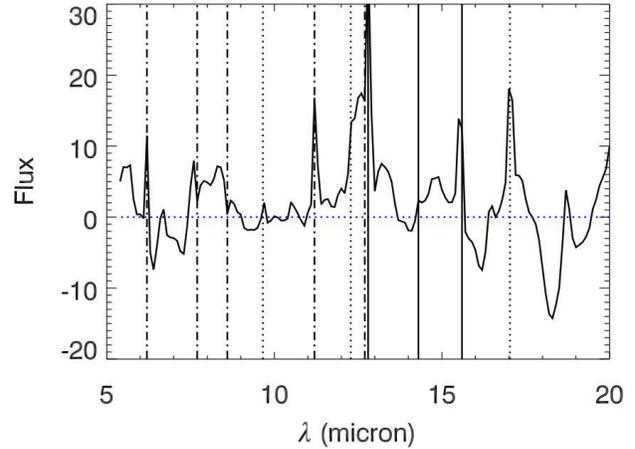}
\caption{The stacked difference between ULIRG spectra and best fit radiative transfer models (solid line) and the ULIRG spectra and 5 PC reconstructions (dotted line). The dot-dashed vertical lines mark the central location of the 6.2,7.7,8.6,11.2 and 12.7 $\mathrm{\mu m}$ PAH emission lines. The dotted lines indicate the location of the molecular hydrogen lines at 9.66, 12.28 and 17.03 $\mathrm{\mu m}$. The solid vertical lines indicate the position of the neon fine-structure lines, [Ne II] 12.8, [Ne v] 14.3 and [Ne III] 15.6 $\mathrm{\mu m}$.}\label{stacked_diff}
\end{figure}

In Figure \ref{stacked_diff}, we show the stacked difference between spectra and radiative transfer model fits (solid line) and the spectra and 5 PC reconstructions (dotted line). The stacked difference for spectra and models illustrates that model fits underestimate the PAH spectral lines and do not include Neon fine structure lines, or molecular Hydrogen lines. The PAH underestimate is consistent with our interpretation of Figure \ref{pcschi}. It suggests the \cite{Kruegel:2003} PAH treatment used by the \cite{Siebenmorgen:2007bc} starburst models, is unsuitable for the extreme starforming ULIRGs. As expected, the PC reconstructions perform considerably better than the models.

\begin{figure}
\includegraphics[width=8.5cm]{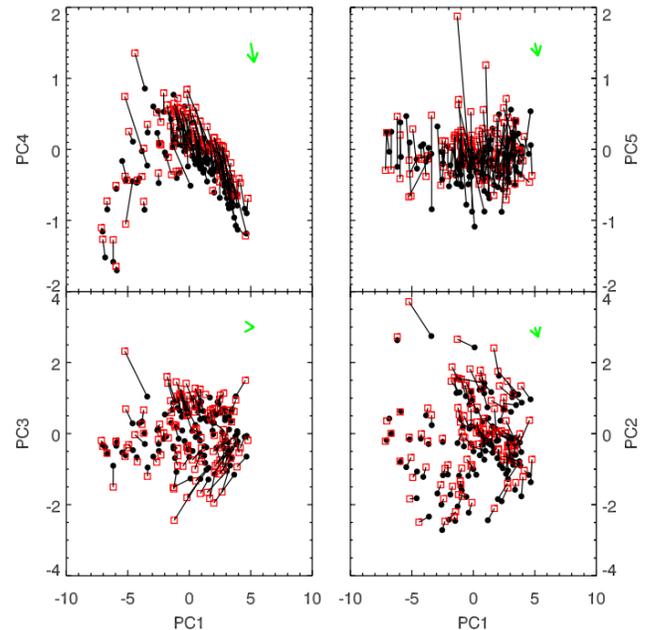}
\caption{The position of ULIRGs in four of the PC planes (squares) and the position in PCA space of the corresponding best fit radiative transfer models (filled circles). Each ULIRG and best fit model are joined by a solid line. The arrows in the top left of each plot show the mean difference between ULIRGs and models.}\label{PCs_PCsmodel}
\end{figure}

\subsection{Interpreting the Principal Components}
\begin{figure}
\includegraphics[width=8.5cm]{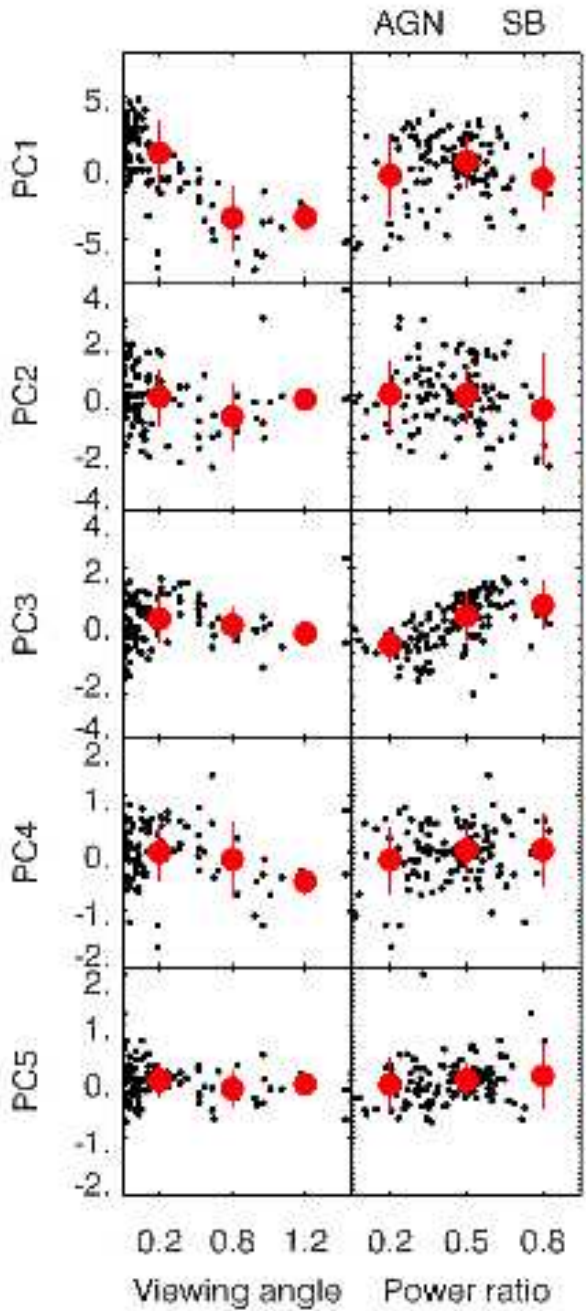}
\caption{The contribution from each PC against the radiative transfer parameters of viewing angle and starburst/AGN contribution. The average contribution for three bins and associated one sigma dispersion are overplotted.}\label{PCs_params}
\end{figure}

\begin{figure}
     \includegraphics[width=8.5cm]{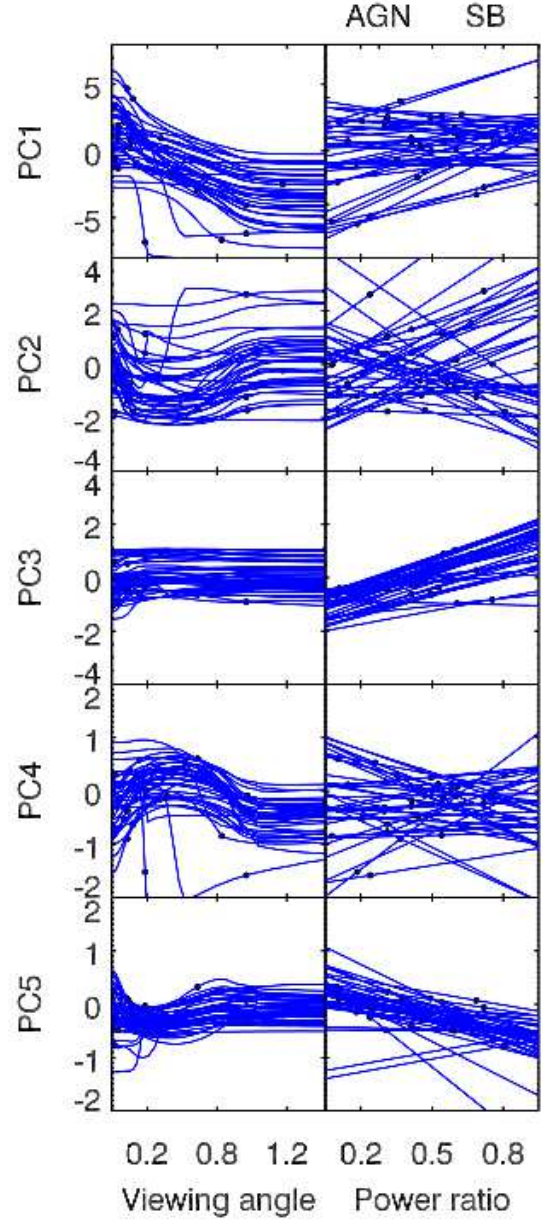}
    \caption{The contribution from each PC against viewing angle tracks and power ratio for 50 of the ULIRGs. For each best fit radiative transfer model, the viewing angle and power ratio have been varied to create tracks in PCA space.}
    \label{tracks}
\end{figure}

We have shown that 5 PCs provide a simple empirical basis set that capture most of the important variations in ULIRGs. We have also shown some limitations of the models. Nevertheless, the models still describe some of the physics of the objects and can be cautiously used to investigate whether the components are associated with physical parameters. We investigate the components by directly comparing the PC contributions and the radiative transfer model best fits for the ULIRG sample. Figure \ref{PCs_params} shows the contribution from each PC as a function of the viewing angle and starburst/AGN contribution. We have binned the PC contributions and calculated the average and one sigma dispersion for each bin. These are over-plotted with errorbars. PC1 shows a correlation with viewing angle of AGN, with positive contributions corresponding to an obscured AGN and negative to face on AGN. The contribution from the fourth PC appears to drop with viewing angle from around $\frac{\pi}{4}$ radians. The other PCs show no discernible dependence. The starburst/AGN contribution is plotted against PCs in the right hand side of Figure \ref{PCs_params}. Negative values of PC3 seem to be associated with AGN dominated sources and positive values with starbursts. The other PCs show a large amount of dispersion and little correlation with starburst/AGN contribution. 

We now decompose the radiative transfer model fits into the PCA space described in Section \ref{PCAspace}. The position of each ULIRG (squares) in four of the PCA planes and corresponding best fit model (filled circles) can be seen in Figure \ref{PCs_PCsmodel}. The mean difference between the ULIRGs and models is depicted by the arrow in the top right of each plane.

We find the location of radiative transfer model best fits in PCA space are offset relative to the ULIRG positions. There are numerous explanations for the offset. The sparseness of the model library could be a factor. The decomposition into PCA space may also be affected by the missing physics in the models. We therefore treat the model tracks with caution and limit interpretation to relative changes in PC contribution rather than absolute position. 

We have taken the best fit radiative transfer model and vary each parameter in turn to see how it affects the position in PCA space. We focus on the viewing angle of AGN and ratio of starburst to AGN power, which we define as:

\begin{equation}
L_{\mathrm{total}}=pL_{\mathrm{SB}}+(1-p) L_{\mathrm{AGN}}
\end{equation}

A value $p=0$ describes a pure AGN model, and $p=1$ relates to a complete starburst.

 Figure \ref{tracks} shows the 1D parameter tracks for 50 randomly selected ULIRGs. The viewing angle tracks show a decrease in PC1 contribution when going from an obscured to face on AGN. Tracks in PC4 are curved, indicating a non-linear relationship with viewing angle. PC3 appears to be a good indicator for the power ratio, with PC3 contribution decreasing as AGN power begins to dominate. Tracks in PC5 also show a slight correlation with power ratio, while for the other PCs the relationship is unclear.

%
%

The interpretation of tracks is consistent with the conclusions drawn from Figure \ref{PCs_params}. Certain PCs appear to be related to the physics of the ULIRGs. We have shown that PC1 is linked to AGN viewing angle, while PC3 is linked to the star formation and AGN contribution. This is consistent with the interpretation of \cite{LingyuPCA}. 
\begin{figure*}
\begin{tabular}{c}
\includegraphics[width=16cm]{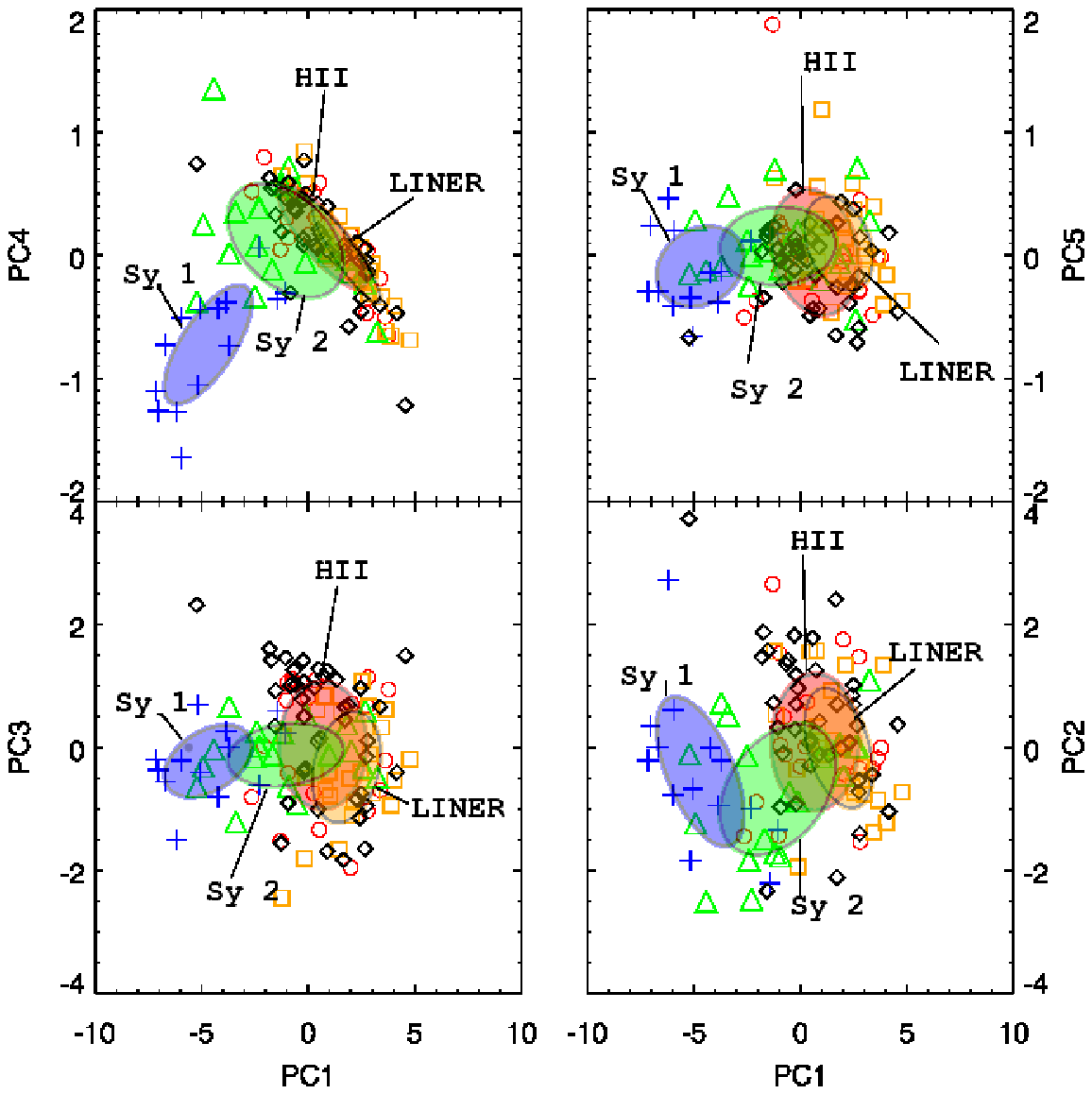}\\
\includegraphics{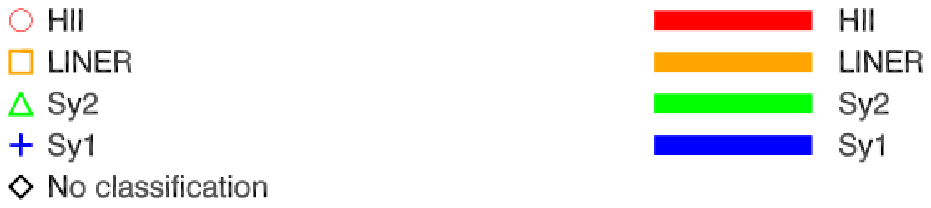}\\
\end{tabular}
\caption{Four out of the possible 10 2D projections for our PCA space with the one sigma contours for the gaussian mixtures based classifications. Optically classified Seyfert 1 objects are marked by crosses, Seyfert 2 by triangles, LINERs by squares and HII classified objects with open circles. Those objects without optical classification are marked by diamonds.}\label{PC1vsPC4}
\end{figure*}

\subsection{Gaussian mixtures classification scheme}
Since we have shown the PCs capture most of the information in IRS spectra, it is natural to use the PCs as a classification tool. \cite{LingyuPCA} suggested that position in the PC1-PC4 plane was related to optical type. We now take take this one step further by proposing a classification scheme based on optical classifications, using the multi-dimensional Gaussian mixtures modelling applied in \cite{Davoodi:2006}. This type of parametric modelling works by assuming the density function of galaxies in our 5D PCA space is composed of a mixture of multidimensional Gaussian functions. We take the four optical classifications (Seyfert 1, Seyfert 2, LINER and HII) that exist for 78 of our 119 ULIRG sample, and assume the density of objects in each classification can be described as Gaussian. The resulting position and width of each Gaussian are trained from the optical classifications. They can be thought of as a probability density function (PDF) that describes the probability of belonging to each optical classification, as a function of position in PCA space.

Figure \ref{PC1vsPC4} shows the marginalised one sigma contours for the optical classifications in four 2D projections. We note that the one sigma contours are for visualisation only, our classification scheme makes use of all 5 dimensions. The objects with optical classifications are represented with different symbols: crosses for Seyfert 1, triangles for Seyfert 2, squares for LINER and open circles for objects classified as HII. Objects without an optical classification are plotted with a diamond. The success rate of our classification, can be found in Table~\ref{tab:optc}.

The classification scheme is very successful in correctly identifying Seyfert 1 like objects, while most of the Seyfert 2s are classified correctly of as LINERs. The majority of LINER objects are correctly identified, while the majority of HII optically classified ULIRGs are spread across HII and LINER groups. Both the LINER and the HII classifications lie in similar areas of PCA space, and discrete classification for objects in this region may not be completely appropriate as many ULIRGs will show signs of both. Overall our 5D Gaussian classification scheme works well in associating regions in PCA space with type of object and is a powerful tool in objectively classifying objects.

\begin{figure*}

\includegraphics[width=16cm]{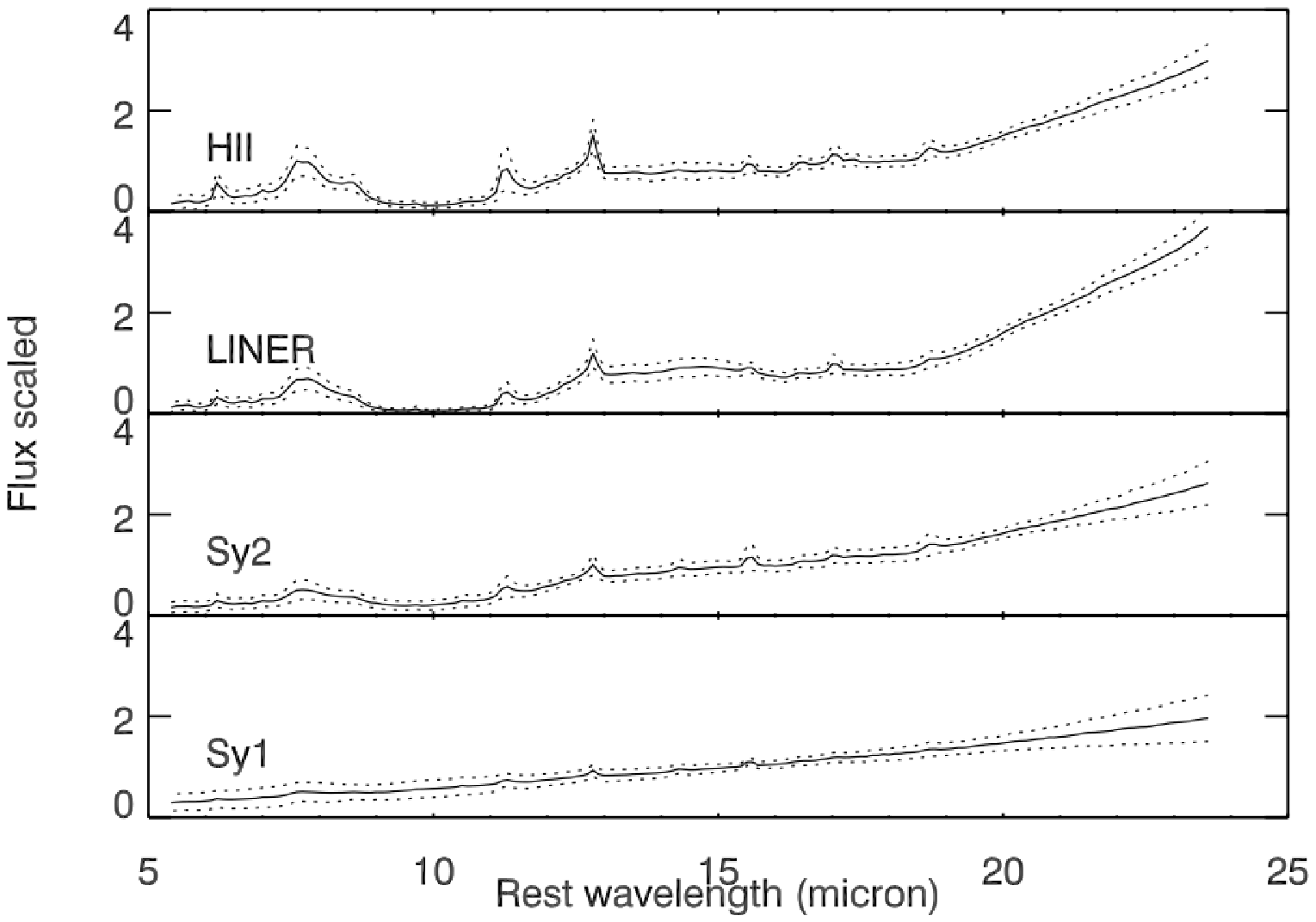}
\caption{The average spectra for the four classifications, HII (29 objects), LINER (54 objects), Seyfert 2 (20 objects) and Seyfert 1 (16 objects). The dotted lines represent the one sigma dispersion in each classification.}\label{stackedspectra}
\end{figure*}

We have used our classification scheme to classify the 41 ULIRGs with no optical classification. The percentages can be seen in Table~\ref{tab:optc}. We find the majority are HII and LINER objects while 12\% are classified as Seyfert 2 like objects. None of the objects appear to be Seyfert 1, suggesting optical classification of Seyfert 1 objects is complete. We now make use of our 5D Gaussian classification scheme by creating average spectra for our four classifications using all 119 ULIRGs. Before averaging the spectra, each spectrum is normalised so that the mean flux over the whole wavelength range is unity. The resulting four average templates can be seen in Figure \ref{stackedspectra}. As expected, the HII and LINER templates are similar, whilst Seyfert templates have very little PAH emission.

\begin{table}
\centering
\begin{tabular}{cc|c|c|c|c|l}
\cline{3-6}
& & \multicolumn{4}{|c|}{Gaussian classification} \\ 
& & HII & LINER & Sy 2 & Sy 1 \\ \cline{1-6}
\multicolumn{1}{|c|}{\multirow{4}{*}{Optical}} &
\multicolumn{1}{|c|}{HII} & 40\% & 45\% & 15\% & 0\%&    \\ 
\multicolumn{1}{|c|}{}                        &
\multicolumn{1}{|c|}{LINER} & 11\% & 85\% & 4\%& 0\%&     \\ 
\multicolumn{1}{|c|}{}                        &
\multicolumn{1}{|c|}{Sy 2} & 5\%& 26\% & 51\% & 18\%& \\ 
\multicolumn{1}{|c|}{}                        &
\multicolumn{1}{|c|}{Sy 1} & 0\% & 0\% & 6\% & 94\% &\\ \cline{1-6}
\multicolumn{2}{|c|}{Not classified} & 42\% & 46\% & 12\% & 0\% &\\ \cline{1-6}

\end{tabular}

\caption{The percentage of objects in the four classifications as a function of their original classification. Not classified refers to those objects without an optical classification.}
\label{tab:optc}
\end{table}

\section{Conclusions}
\label{conclusions}
We have shown that five principal components are needed to describe most of the variation in the 119 local ULIRG sample and are more successful than a full $\chi^{2}$ fitting by radiative transfer models. We have examined what the radiative transfer models are missing. The fits provided by radiative transfer models appear to need a cold dust component and have difficulty in modelling the strength of strong PAH emission lines.

We have used a combination of the \cite{Siebenmorgen:2007bc} starburst models and \cite{Efstathiou:1995ge} AGN torus templates to investigate what physical parameters are behind the components. We have examined how best fit model parameters are related to PC contribution. Overall, our conclusions are consistent with those reached in \cite{LingyuPCA}. Contributions from PC1 appear to indicate the viewing angle of AGN with negative contributions associated with face on AGN and positive for obscured AGN. PC3 appears to be the best indicator of whether it is the AGN or starburst that is the prevailing power source. 

The PCs consider a large part of the mid-infrared spectrum and are therefore less likely to be affected by problems associated with diagnostics based on single spectral features such as the PAH emission lines, where measuring line strength can be difficult. We suggest the five PCs would be useful as empirical templates for ULIRG spectra in the IRS public database \citep{Lebouteiller:2011}. 

We also introduce a new Gaussian mixtures classification scheme based on location in the five dimensional PCA space and trained via optical classifications. Objects can be classified as either Seyfert 1, Seyfert 2, LINER or HII-like. We note that any ULIRG with IRS spectra (in the relevant wavelength range) can be decomposed onto the PCs, and the position in PCA space can be used to classify the object. 

We have used our classification scheme to provide a set of average spectra for the four groups. We make these, the five PCs and code to classify objects available at: \url{http://www.phys.susx.ac.uk/~pdh21/PCA/}

\section*{Acknowledgements}
We acknowledge support from the Science and Technology Facilities Council [grant numbers ST/F006977/1, ST/I000976/1, PP/E005306/1]. This work is based on observations made with the Spitzer Space Telescope, which is operated by the Jet Propulsion Laboratory, California Institute of Technology under a contract with NASA. We thank the referee for the very helpful comments which led to improvements in this paper. 
\bibliography{sortedbib050710.bib}
%
%
%
%
%
%
%
%

\end{document}